\pdfoutput=1
\documentclass[11pt,oneside,hidelinks]{article}
\usepackage{geometry,amsmath}
\usepackage[tight,nice]{units}
\usepackage[numbers]{natbib}
\usepackage{hyperref}
\hypersetup{
  colorlinks   = true, 
  urlcolor     = blue, 
  linkcolor    = blue, 
  citecolor   = red 
}
\usepackage{graphicx,float,amssymb}
\usepackage{graphicx}
\usepackage{wrapfig}
\usepackage[font=small,labelfont=bf]{caption}
\usepackage{csvsimple}
\usepackage{tabularx}
\usepackage{ragged2e}
\usepackage{physics}
\usepackage{subcaption}
\usepackage[usenames, dvipsnames]{color}

\newcommand {\be}{\begin{equation}}
\newcommand {\ee}{\end{equation}}
\newcommand {\ba}{\begin{eqnarray}}
\newcommand {\ea}{\end{eqnarray}}
\newcommand {\bal}{\begin{aligned}}
\newcommand {\eal}{\end{aligned}}

\title{Quantum phase transition modulation in an atomtronic Mott switch}

\begin{document}

\maketitle

\begin{centering}

Marie A. McLain and Lincoln D. Carr

\textit{Department of Physics, Colorado School of Mines, Golden, CO, USA}

\end{centering} 

\section{Abstract}
Mott insulators provide stable quantum states and long coherence times due to small number fluctuations, making them good candidates for quantum memory and atomic circuits. We propose a proof-of-principle for a 1D Mott switch using an ultracold Bose gas and optical lattice. With time-evolving block decimation simulations -- efficient matrix product state methods -- we design a means for transient parameter characterization via a local excitation for ease of engineering into more complex atomtronics. We perform the switch operation by tuning the intensity of the optical lattice, and thus the interaction strength through a conductance transition due to the confined modifications of the ``wedding cake'' Mott structure. We demonstrate the time-dependence of Fock state transmission and fidelity of the excitation as a means of tuning up the device in a double well and as a measure of noise performance. Two-point correlations via the $g^{(2)}$ measure provide additional information regarding superfluid fragments on the Mott insulating background due to the confinement of the potential.

\section{Introduction}
\label{sec:phintro}
Ultracold bosons in optical lattices provide highly precise architectures for quantum simulation of systems from solid state materials and superconductors to nonlinear optics. Due to their tunability, parameters governing cold atom dynamics such as interactions, temperature, and defect formations are highly controllable in arbitrary ``painted'' potentials~\cite{henderson2009,islam2015,Zhao2017}. Atomtronics -- the creation of atomic circuits analogous to their electronic counterparts -- is an emerging field in optical lattice platforms because it provides direct translations to descriptions such as solid state circuit elements and batteries~\cite{Balachandran2017,seaman2007,li2016,lai2016,caliga2017}, interferometers~\cite{Haug2017}, transistors~\cite{pepino2009,caliga2016,Vaishnav2008}, superconducting or atomtronic quantum interference devices~\cite{ryu2013,mathey2016,haug2018,eckel2014}, and even open quantum circuits~\cite{lai2016,caliga2017,amico2017}. By probing the influence of the Mott phase transition on insulator conductance properties, we aim to demonstrate the opportunity for multi-disciplinary simulations of both semi-classical and quantum systems. The technology readiness level has matured to the point that there is a new need for quantum engineering; this has been achieved through advancements both in computation with numerical methods, in particular matrix product states~\cite{Scholl2011,ref:OSMPS} -- well-suited for strongly-interacting 1D lattice chains -- and in experiment with Feshbach resonances, trapping techniques, and site-resolved imaging~\cite{shuman2010,TGgas2006,greif2016,feshbach,parsons2015,haug2018}. More fully-quantum descriptions such as those offered by matrix product state methods such as time-evolving block decimation can offer insights beyond semiclassical atomtronic descriptions~\cite{olsen2015,draxler2017,jaschke2018}.

Additionally, Bose-Einstein condensate interferometry experiments in multiple potential wells admit dephasing is the largest source of error and limits the noise floor on coherence times~\cite{burton2016}.  Mott insulating atomtronics have the advantage of low number fluctuations and longer coherence times as compared to their weakly-interacting counterparts. We build on initial Mott atomtronic circuits~\cite{pepino2009} by presenting a proof-of-principle of a single Mott-insulating switch with a modulation tunable via a quantum phase transition, and a transient analysis method distinct from forward or reverse biasing methods~\cite{seaman2007}. A switch is one of the simplest devices in a circuit; fundamentally, it can only have two states. Switches provide the backbone for Boolean logic and comprise the heart of classical computation through transistors: they are a foundational circuit element that need to be incorporated into larger Mott insulating circuits in order for atomtronics to expand in scope. We test the otherwise passive circuit element through a transient analysis in response to an impulse and compare the junction leakage to the transmission tunneling coefficient. We use time-evolving block decimation (TEBD) simulations to harness local measures, such as experimentally-observable number density, along with entropy measures to highlight the implications of entanglement in determining atomic conductance in a bosonic Mott insulator. To characterize noise in the system, we observe a microscopic measure -- the Fock state fidelity -- which offers a clean time domain in the operating regimes, and thus, a discrete Fourier transform of the fidelity offers a robust metric of the noise floor. In addition, the $g^{(2)}$ correlation measure enables many-body characterization that is directly translatable to experiment~\cite{Song2010}. We use $g^{(2)}$ to quantify superfluid fragmentation; thus we are able to isolate the more pure Mott insulator that is the switch in its normal operation from these superfluid fragments that manifest on top of the Mott background, signaling the disconnected switch state.

\begin{figure}
\centering
\includegraphics[width=0.86\textwidth]{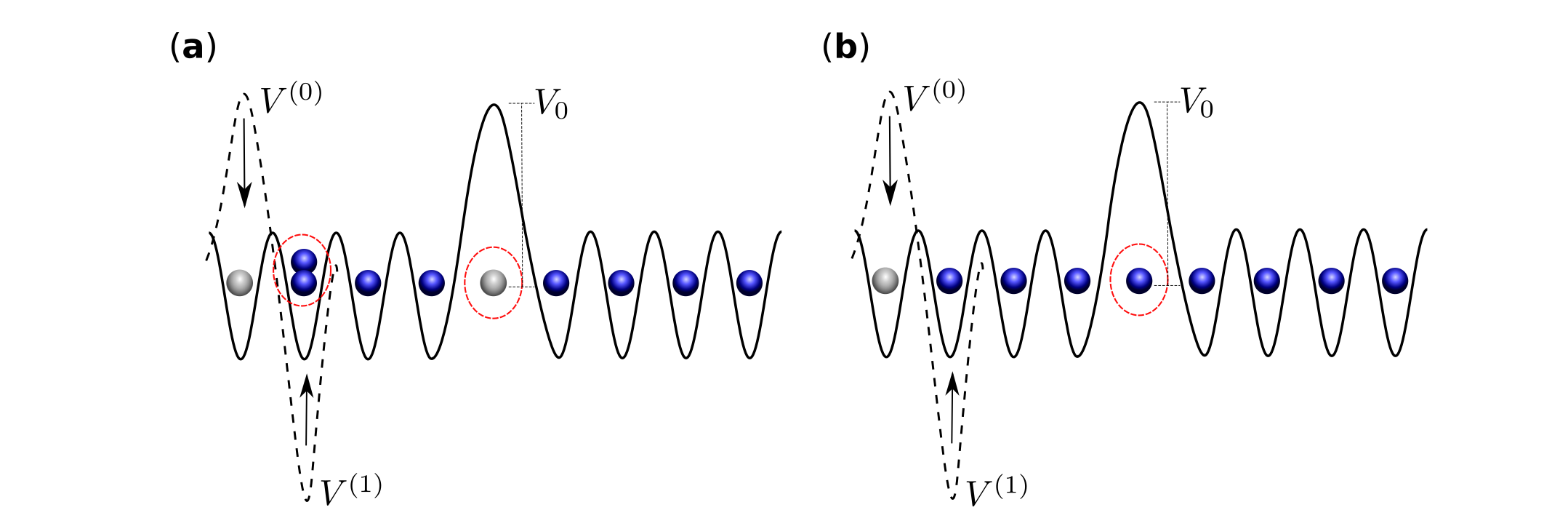}
\caption[Transient source quench scheme.]{\label{fig:cartoonSwitch}\textit{Transient source quench scheme.} We initialize a Mott insulator in a bosonic Mott-insulating junction with a localized excitation, shown as the dashed lines of the potential, by running imaginary time propagation. The local kick in the first two sites of the potential is diabatically quenched, as indicated by the arrows, to the uniform lattice depth indicated with solid curves. The excitation allows for transient characterization of the switch parameters. (\textbf{a}) Below an interaction threshold, a particle-hole excitation forms in the initial state that has low conductance. (\textbf{b}) Above the interaction threshold, the extra boson from the particle-hole prefers to occupy the barrier region, and the excitation instead instigates a hole in the initial state with high conductance properties.}
\end{figure}

We design the switch as a double well potential with an underlying optical lattice, as shown schematically in Figure~\ref{fig:cartoonSwitch}. The 1D potential is well-described by the Bose-Hubbard Hamiltonian,

\begin{equation}
\label{eqn:BHHswitch}
\hat{H}_{\mathrm{BH}} = {} -J\sum_{\langle i,j\rangle}(\hat{b}^{\dagger}_i \hat{b}_{j}+\hat{b}_i \hat{b}^{\dagger}_{j}) +
\frac{1}{2}U\sum_i \hat{n}_i(\hat{n}_i-1) + 
\sum_i V_i\hat{n}_i\,,
\end{equation}
such that $J$ is the bosonic tunneling energy, $U$ is the repulsive interaction energy, $\hat{b}^{\dagger}_i$ and $\hat{b}_i$ are bosonic creation and annihilation operators, respectively, which satisfy bosonic commutation relations, $\hat{n}_i$ is the number operator for bosons, $\langle i,j\rangle$ indicates a sum over nearest neighbors, and the indices $i,j \epsilon \{0,...,L-1\}$ are over the $L$-length of the 1D optical lattice. For the purposes of modulating lattice depth in this study, we take $J=1$ in all cases. $V_i$ is the confining potential of the double well, and we will refer to $V_0$ as the central barrier height.

To initialize a transient response, we use imaginary time propagation in TEBD to find a modified Mott ground state to the dashed potential in Figure~\ref{fig:cartoonSwitch}, where a majority of the lattice is in a unit-filled Mott state, i.e. an average of one atom per site. The first two sites have an initial excitation induced from the potential, which can be created e.g. with lasers as in atom gas microscopes~\cite{yamamoto2016,sherson2010,Bakr2010}. Due to this local excitation and the commensurate filling of bosons on the lattice, the first site has a hole: it is void of any particles. However, the strength of the interaction energy $U$ and the potential barrier -- which has the width of a single lattice site -- compete to determine whether the local excitation is a simple hole in the Mott background (Figure~\ref{fig:cartoonSwitch}(\textbf{b})) or whether it is a particle-hole pair (Figure~\ref{fig:cartoonSwitch}(\textbf{a})). As we explore further in the results, there is a critical value of interaction strength that dictates whether the excitation is a hole or a particle-hole pair, and this same critical value also governs the conductance of the excitation. Of course, for shallow enough lattice depths, there is an additional threshold where tunneling dominates, so particle-hole creation is no longer energetically favorable. This second threshold is below the Mott limit in one dimension, $U/J \lesssim 3$~\cite{Cristiani2002}. Fundamentally, it is the presence of the barrier that changes the usual Mott ground state enough to let a hole located at the barrier be preferred over a particle-hole. This is similar to changes in a harmonic trap resulting in a wedding-cake structure.  Once the initial state has been established, we diabatically quench the first two sites so the underlying optical lattice is uniform, and we measure on-site particle number and fluctuation dynamics. From the local occupation, we calculate reflection and transmission probabilities of the hole through the double well barrier as a function of time, with oscillations indicating the normal operation of the switch and self-trapping indicating the switch is disconnected.

\section{Results and Discussion}
\label{sec:phresults}

\begin{figure}
\centering
\includegraphics[width=0.85\textwidth]{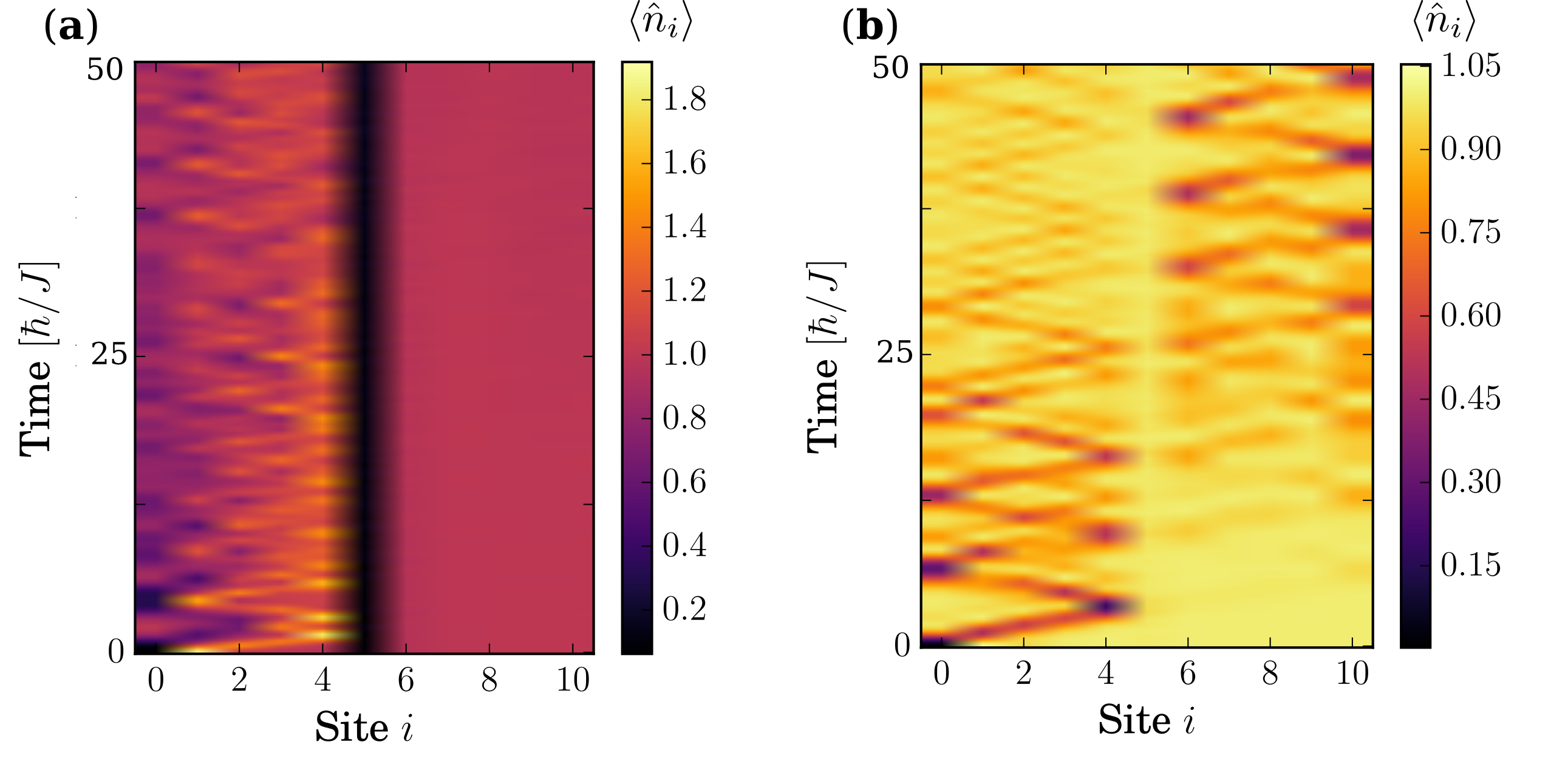}
\caption[Interaction switch modulation due to ``wedding cake'' Mott insulator.]{\label{fig:cakeSwitch}\textit{Interaction switch modulation due to ``wedding cake'' Mott insulator.} The two possible transient response states of the classical Mott switch are (\textbf{a}) self-trapping in the left well and (\textbf{b}) Josephson oscillations. The barrier height $V_0=5$, in both panels, is a sweet spot for switch operation in this system configuration as determined by the number of lattice sites, the interaction strengths, the excitation magnitudes, and the atomic filling factor. The critical switching manifests as a sharp transition between (\textbf{a}) $U=25$ and (\textbf{b}) $U=26$, with all other parameters held constant, indicating the phase transition rather than tunneling is responsible for the division.}
\end{figure}

One method for switch modulation is detailed in~\cite{mclain2018} that dictates the dynamic regimes of a bosonic Josephson junction in a 1D optical lattice based on height of the central potential barrier. The states of the switch correspond to the two regimes of the spontaneous-symmetry breaking $\mathbb{Z}_2$ quantum phase transition in a 1D double well: Josephson and Fock. In the single-particle limit, the Josephson oscillations in an isolated bosonic Josephson junction approach the Rabi frequency of the double well. As the barrier height is increased beyond some threshold, the junction enters the Fock regime and bosons remain self-trapped on one side of the barrier, an effect also observed as a Coulomb blockade in superconducting Josephson junctions~\cite{Haviland1991}. Signatures of this quantum phase transition persist and dominate the dynamics even for mesoscopic systems with few active degrees of freedom. The results in Figure~\ref{fig:cakeSwitch}(\textbf{a}) can be interpreted as analogous, where the Fock regime indicates the switch is disconnected and Figure~\ref{fig:cakeSwitch}(\textbf{b}) portrays the Josephson regime, where the switch is connected. The key is that the typical $\mathbb{Z}_2$ shift from macroscopic quantum self-trapping to the Josephson regime as dictated by tunneling through the barrier is smooth and continuous~\cite{lcarr2010,Sachdev2011} -- whereas the phase transition assisted switch in this article exhibits sharp contrast between the switching states. In our case we make use of the strongly-correlated Mott insulating state, leading to a more robust switch and much more localized single-site excitations, allowing for ultimately smaller atomtronic elements.

The switch modulation method we investigate in this article relies on the well-known ``wedding cake'' model of the Mott insulator to superfluid transition in the presence of a trapping potential~\cite{batrouni2002}. Unconfined optical lattice models of the Mott insulator, for instance, require particle numbers commensurate with the number of lattice sites. The confining potential changes this definition of commensurate fillings, and thus the particle numbers needed to achieve a Mott insulator may be slightly different from traditional unit fillings. This is the case in our present study, where we have 11 lattice sites, 5 in each well, with the barrier occupying a single lattice site. We find a proper filling factor using 10 bosons in this Mott insulator. While we focus on the 1D chain of 11 lattice sites for the data in this article, we find that in order to employ the switch modulation method we propose, for a system of $L$ lattice sites the number of total atoms must be $N = n(L-1)$, where $n$ is an integer dictating the commensurate filling factor. Though, it is no doubt possible for other designer lattice topologies and sizes to exploit similar phase transition induced control, for instance generalizing to larger barriers, in which case the condition would be $N = n(L-w)$, with $w$ the barrier width. To determine experimental feasibility without the requirement for single-site control, we systematically confirmed the effect for barriers up to a width of 3 lattice sites for total lattice sites $L=9,11,13,15$ from $U=4$ through $60$ where we are in the Mott insulating regime. We consider odd numbers of lattice sites because with a barrier width of 1, this yields an even number of sites on either side to maintain symmetry. For a barrier width of 2,4, etc. it is appropriate to have an even number of lattice sites. Thus, because we find the effect robust for varying barriers, the particle number would scale with the number of lattice sites as would be expected for a unit-filling Mott insulator. While the experimental control of a precise particle number presents a challenge, the minimum experimental implementation would require a Mott insulator with the presence of a double well confining potential, where the number of holes in the Mott background due to the confinement of the potential can indeed vary. Advances in experimental precision may also help to realize the Mott switch, such as controllable two-body collisions for finer Mott manipulation~\cite{Tomita2017}. The critical point occurs between $U=20$ and $30$ for barrier heights $V_0=0.1$ to $10$ for $L=11$ sites and $N=10$ bosons, as will be discussed throughout this article. We intentially focus on small systems to minimize the size of the atomtronic device, working at the smallest size for which a quantum phase transition is still a relevant concept, about 10 sites plus the barrier accounting for finite size effects~\cite{CarrWall2010}.

\begin{figure}
\centering
\includegraphics[width=0.99\textwidth]{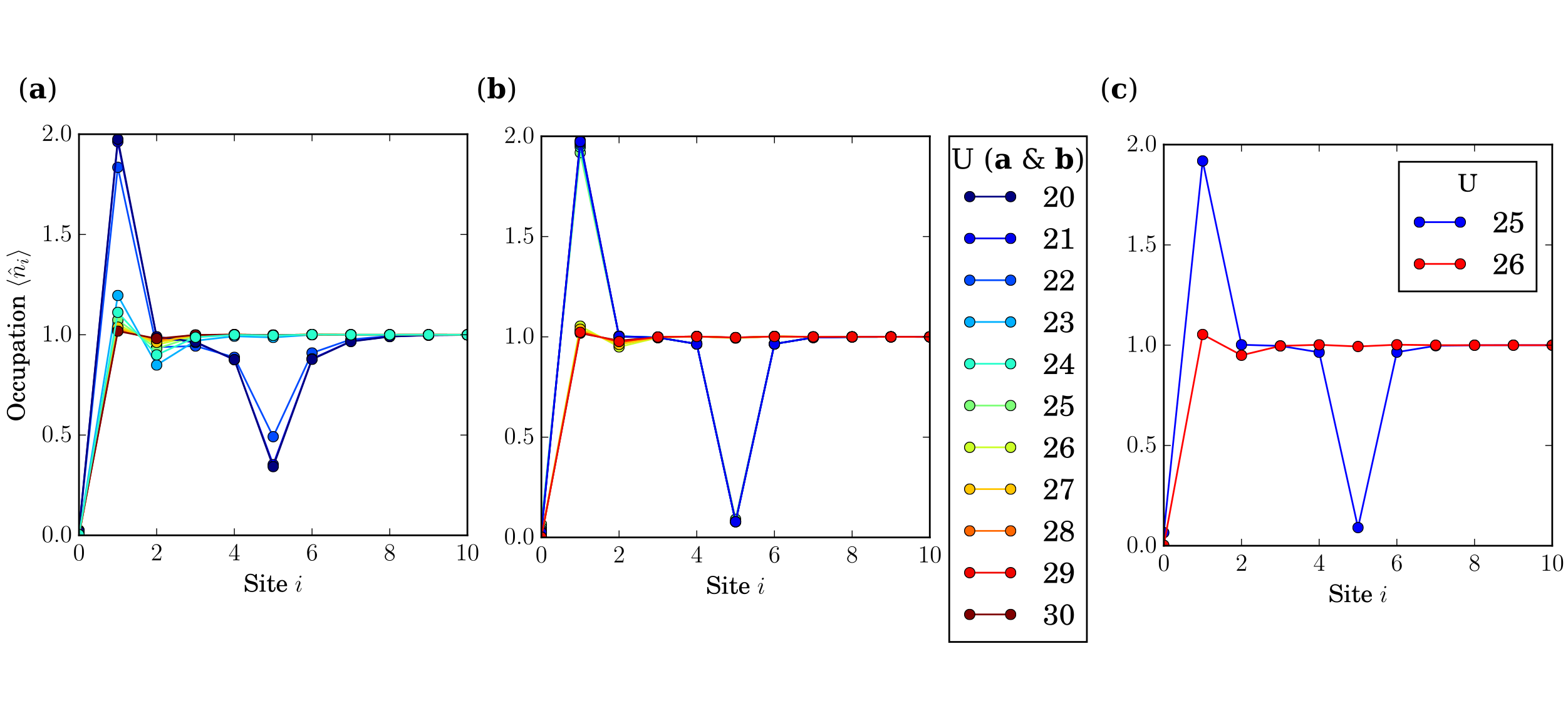}
\caption[Initial states suggest critical Mott confinement.]{\label{fig:switchInitSts}\textit{Initial states suggest critical Mott confinement.} (\textbf{a}) When the barrier height $V_0=2$, a particle-hole pair forms on the first two sites due to the induced excitation of the initial states for $U=20$ and $21$. For larger interaction strengths, the extra boson occupies the central barrier rather than the particle-hole pair. (\textbf{b}) At $V_0=5$, the initial states approach ideal Mott states as the on-site occupations converge toward integer values. (\textbf{c}) The critical transition from particle-hole excitation to barrier occupation occurs between $U=25$ and $26$ for $V_0=5$.}
\end{figure}

We set the tunneling strength $J=1$, the central barrier height $V_0=5$, and the initial excitation strengths on the zeroth site $V^{(0)}=10$ and first site $V^{(1)}=-10$, which provide a localized kick that is strong compared with the barrier height. While these excitation strengths were originally designed to favor particle-hole formation on the first two sites, we find that the critical transition to barrier occupation occurs, surprisingly, in spite of the imbalance of the positive and negative excitations. While these excitations can be incited at any site(s) on the lattice, in order to control the direction of propagation we initialize it adjacent to the leftmost wall, which we note has open boundary conditions. In Figure~\ref{fig:switchInitSts}(\textbf{a}), the interaction strength $U=25$ is below its critical value for turning on the switch, and thus the particle-hole is self-trapped in the left well and the right well remains in an ideal Mott state. The switch is off, or in its disconnected state. The initial state has close to no particle occupation in the barrier, whereas Figure~\ref{fig:switchInitSts}(\textbf{b}) with $U=26$ illustrates the repulsive interaction is critically strong enough such that the particle prefers to occupy the barrier, thus creating a conductive link to the right well. The switch is therefore on, or in its normally operating state.

An important clue in deciphering the influence of the superfluid-Mott phase transition on the operation of this Mott device is analyzing the initial states. We supply a pure Fock state as an initial guess, $| 0 2 1 1 1 0 1 1 1 1 1 \rangle$, into our TEBD imaginary time propagation algorithm, though ultimately the ground state determination of this initial potential has no absolute requirement for initial Fock states. This can be seen in Figure~\ref{fig:switchInitSts}(\textbf{a}) for $V_0=2$, as the barrier regions on the $5^{\mathrm{th}}$ lattice site for $U=20$ and $21$ are closer to $\langle \hat{n}_5 \rangle \approx 0.5$ than an integer. These deviations from Fock states are due to superfluid influences on the edge of one layer of the Mott insulating shelf; as a contrast, interaction strengths $U>21$ are firmly rooted on the Mott side of the transition, with close to unit filling. Additionally, determining the best confinement parameter will optimize the contrast and effectiveness of the switch. In Figure~\ref{fig:switchInitSts}(\textbf{b}) we see that a barrier height of $V_0=5$ enables more ideal Fock states and thus, less error in the on-site number as a measure of the switch state. As shown in Figure~\ref{fig:switchInitSts}(\textbf{c}), the values for $U\geq 26$ at this barrier height are nearly identical $| 0 1 1 1 1 1 1 1 1 1 1 \rangle$, at unit filling save for the hole on the first site. For $U \leq 25$ the initial states approach the ansatz $| 0 2 1 1 1 0 1 1 1 1 1 \rangle$, and this hole in the barrier region creates a large resistance and inhibits the particle-hole transport. We note that the particle number is conserved in our model, as appropriate for an ultracold Bose gas in an optical lattice.

\begin{figure}
\centering
\includegraphics[width=0.9\textwidth]{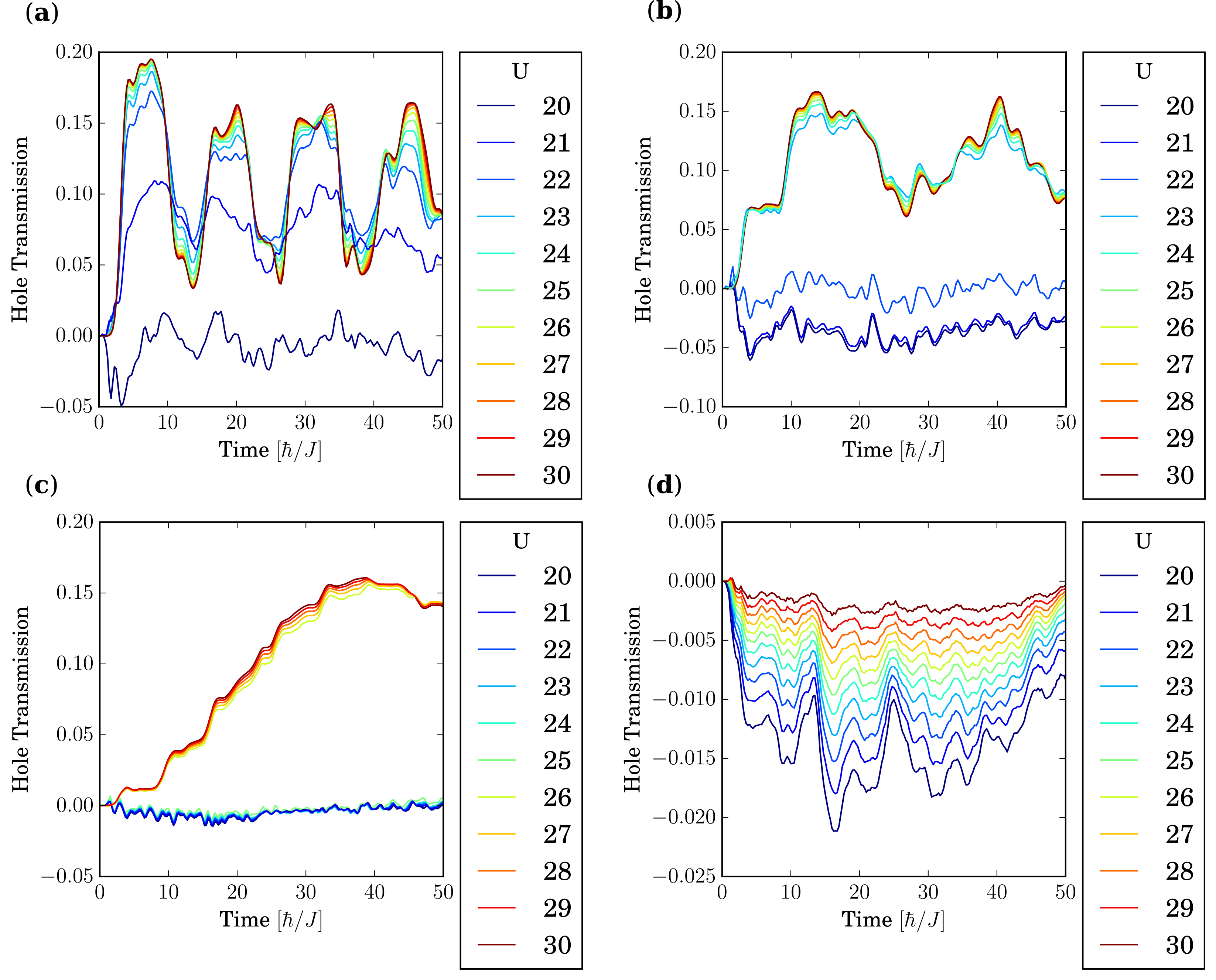}
\caption[Transmission demonstrates regime with best signal-to-noise ratio.]{\label{fig:switchTransmission}\textit{Transmission demonstrates regime with best signal-to-noise ratio.} (\textbf{a}) In the low barrier height regime, $V_0=0.1$, the hole transmits easily for interaction strengths $U\geq 22$. We begin falling off the shelf of the Mott wedding cake for $U=21$ and $20$ as the superfluid washes out the excitation, and the signal-to-noise is low, $\approx 1.6$. (\textbf{b}) For $V_0=2$, the critical split in transmission behaviors between the two possible types of initial states, occurring between $U=22$ and $23$, becomes exaggerated. However, the the signal-to-noise is still low at $\approx 3.4$. (\textbf{c}) The divergence in behavior of the two switching states becomes clear for a barrier height $V_0=5$, and the signal-to-noise ratio is maximal at $\approx 8.9$. ({\textbf{d}}) As the barrier height becomes too large, the transmission of the hole excitation is negligible, and the only transmissions through the barrier are those due to superfluid influences as a result of optical lattice depth.}
\end{figure}

In investigating the response of the device to the sudden excitation, we consider a Fock transmission of the hole across the barrier region based on inverse number density, as the hole is an absence of a particle within the Mott insulating background. We calculate the time-dependent transmission as the dimensionless ratio

\be
\label{eqn:transmission}
T(t) = \frac{\sum_{i=1}^{\left \lfloor{L/2}\right \rfloor} \hat{n}_i(t) - \sum_{i=1}^{\left \lfloor{L/2}\right \rfloor} \hat{n}_i(t=0)}{N/2}, 
\ee
where $L$ is the total number of lattice sites and $N$ the total number of atoms. The sum is over the left well sites $i$, where we could have easily calculated the inverse quantity for the right well, as the double well conserves particle number. The transmission provides one means of measuring the signal-to-noise ratio, which we determine by taking the ratio of the maximum double well oscillation amplitude to the amplitude of the junction leakage, or the magnitude of the maximum noise when the switch is off. We calculate the signal-to-noise based on the integer values of $U$ immediately on either side of the switching phase transition, e.g. $U_{\mathrm{on}}$ ($U_{\mathrm{off}}$) for the interaction strength when the switch is on (off). The signal is then $(\mathrm{max}(|T|) - \mathrm{min}(|T|))\bigg\rvert_{U_{\mathrm{on}}}$ and the noise is $(\mathrm{max}(|T|) - \mathrm{min}(|T|))\bigg\rvert_{U_{\mathrm{off}}}$. In Figure~\ref{fig:switchTransmission}(\textbf{a}) at a low barrier height of $V_0=0.1$, the shift from off to on occurs from $U=21$ to $U=22$ with a small signal-to-noise ratio of $\approx 1.6$, where the error on the measurement is machine error and is not represented in the truncation of the decimal. The next barrier height, $V_0=2$ displayed in Figure~\ref{fig:switchTransmission}(\textbf{b}),  depicts a mildly improved signal-to-noise of $\approx 3.4$ between the normally operating tunnel state at $U=23$ and the disconnected self-trapping at $U=22$. This operation region of the switch is an improvement over the $V_0=0.1$ case, as the phase transition occurs more sharply through the interaction modulation. The switch can be tuned to a large signal-to-noise ratio of $8.9$ between the off state at $U=25$ and the on state at $U=26$ for a barrier height of $V_0=5$ as shown in Figure~\ref{fig:switchTransmission}(\textbf{c}), providing a robust switching configuration that also offers the most contrast between the two states. The final panel of Figure~\ref{fig:switchTransmission}(\textbf{d}) displays self-trapping for all $U$ and the switch is disconnected. The background loss increases for decreasing interaction strength, as the superfluid fraction grows, thus mediating tunneling through the barrier.

\begin{figure}
\centering
\includegraphics[width=0.9\textwidth]{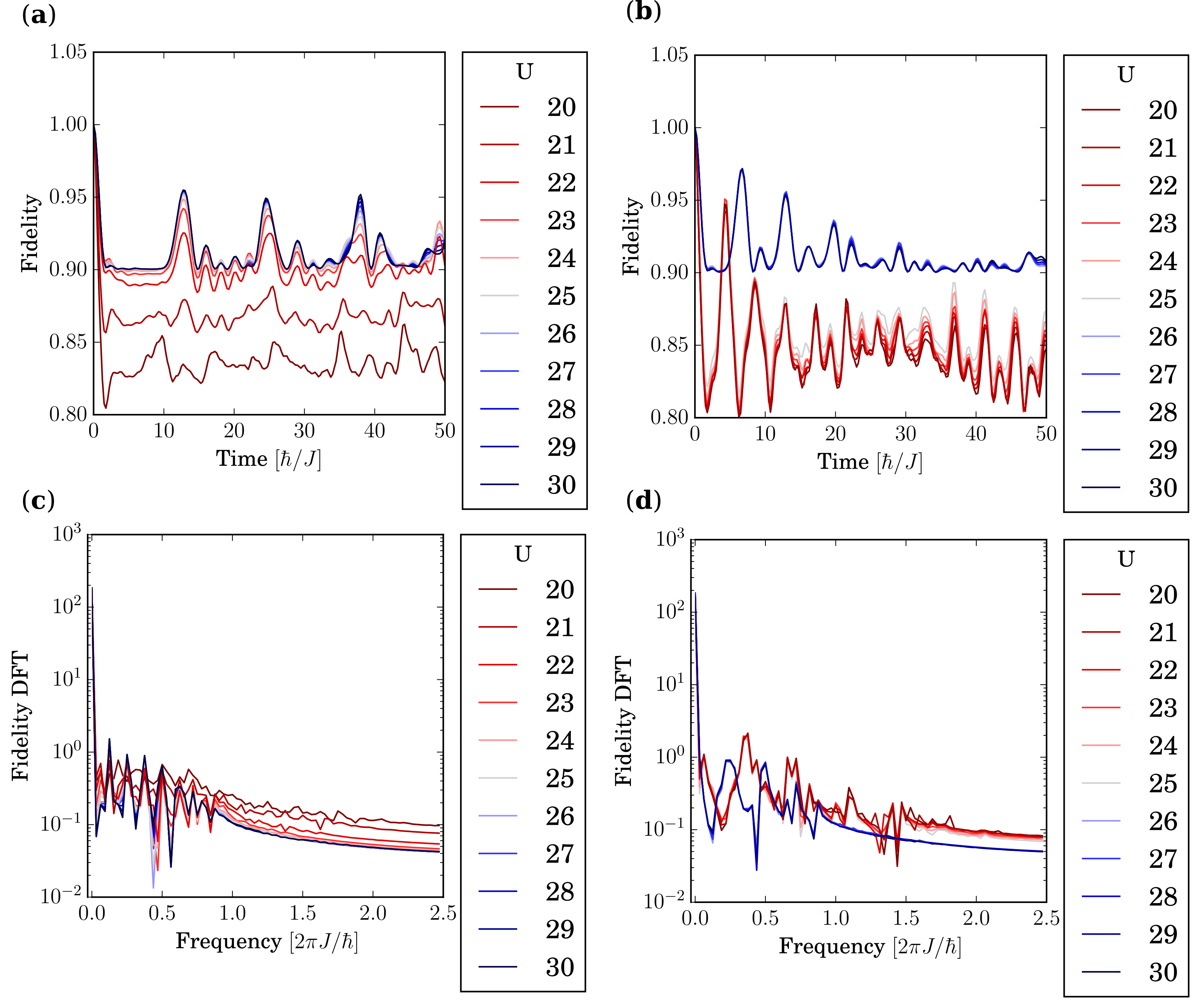}
\caption[Fidelity and corresponding Fourier transform as robust measurement standards.]{\label{fig:switchFidelity}\textit{Fidelity and corresponding Fourier transform as robust measurement standards.} (\textbf{a}) In the low barrier height regime $V_0=0.1$, where $U\geq 22$, the fidelity has a lower limit of $\approx 90 \%$ and exhibits metronomic periodicity. This barrier height is too low to observe a critical transition necessary for a switch. (\textbf{b}) The two states of the Mott switch converge for $V_0=5$, where $U=20$ to $25$ in red means the state of the switch is self-trapped or disconnected, and the initial states were all particle-hole excitations. Interaction strengths $U = 26$ to $30$ in blue illustrate the Josephson or normally operating regime, and the initial states all demonstrated particle occupation of the barrier. (\textbf{c}) The discrete Fourier transform of the data from (\textbf{a}) reveals the continuous raising of the noise floor by approximately 6 dB for an increasing superfluid fraction, as interaction strength decreases from $U=30$ to $20$. (\textbf{d}) The Fourier transform of the data from (\textbf{b}) reveals a divergence in the spectral behavior between the two switching states and the emergence of higher-frequency superfluid noise when the switch is off for $U \leq 25$.}
\end{figure}

The fidelity provides a microscopic measure of transient response in atomtronic Mott insulators. We calculate the fidelity as a Fock space overlap with the initial state, which is $\langle n(0) | n(t) \rangle/ \langle n(0) | n(0) \rangle$, where $| n(t) \rangle$ is the time-dependent Fock state and $| n(0) \rangle$ is the initial Fock state prior to real time propagation. The clean periodic fidelity peaks in Figure~\ref{fig:switchFidelity}(\textbf{a}) are more easily mapped to their Fourier transform, panel (\textbf{c}), than transmission, for example. Below the critical point, $U < 22$, the superfluid fragments as portrayed in Figure~\ref{fig:switchFidelity}(\textbf{a}) lower the fidelity below the $\approx 90 \%$ Mott threshold and in Figure~\ref{fig:switchFidelity}(\textbf{c}) inject high frequency noise into the discrete Fourier transform, calculated using the one-dimensional fast Fourier transform algorithm in numpy. The superfluid influence also raises the noise floor by about 6 dB. Figure~\ref{fig:switchFidelity}(\textbf{b}) displays a better operating regime for the switch with a barrier height $V_0=5$, as the contrast between the two switching states is maximized, such that $U \geq 26$ in blue, the switch is in its normally operating, or on state and $U \leq 25$ in red, the switch is disconnected, or off. The phase transition through $U$ is sharp, yet the operating parameter regions are large, making this switch practical for scalable Mott atomtronics. In addition, Figure~\ref{fig:switchFidelity}(\textbf{d}), the Fourier transform of Figure~\ref{fig:switchFidelity}(\textbf{b}), demonstrates two distinct spectra, the blue for larger $U$ corresponding to Josephson oscillations when the switch is on and the red for smaller $U$ when the switch is off. In this off state, the noise floor is considerably higher by about 3 dB. Also in the off state, the fidelity oscillations are well-resolved for the particle-hole excitation bouncing back and forth within the left well, lending to a more well-defined spectrum for a range of interaction strengths as compared with the $V_0=0.1$ case.

\begin{figure}
\centering
\includegraphics[width=0.9\textwidth]{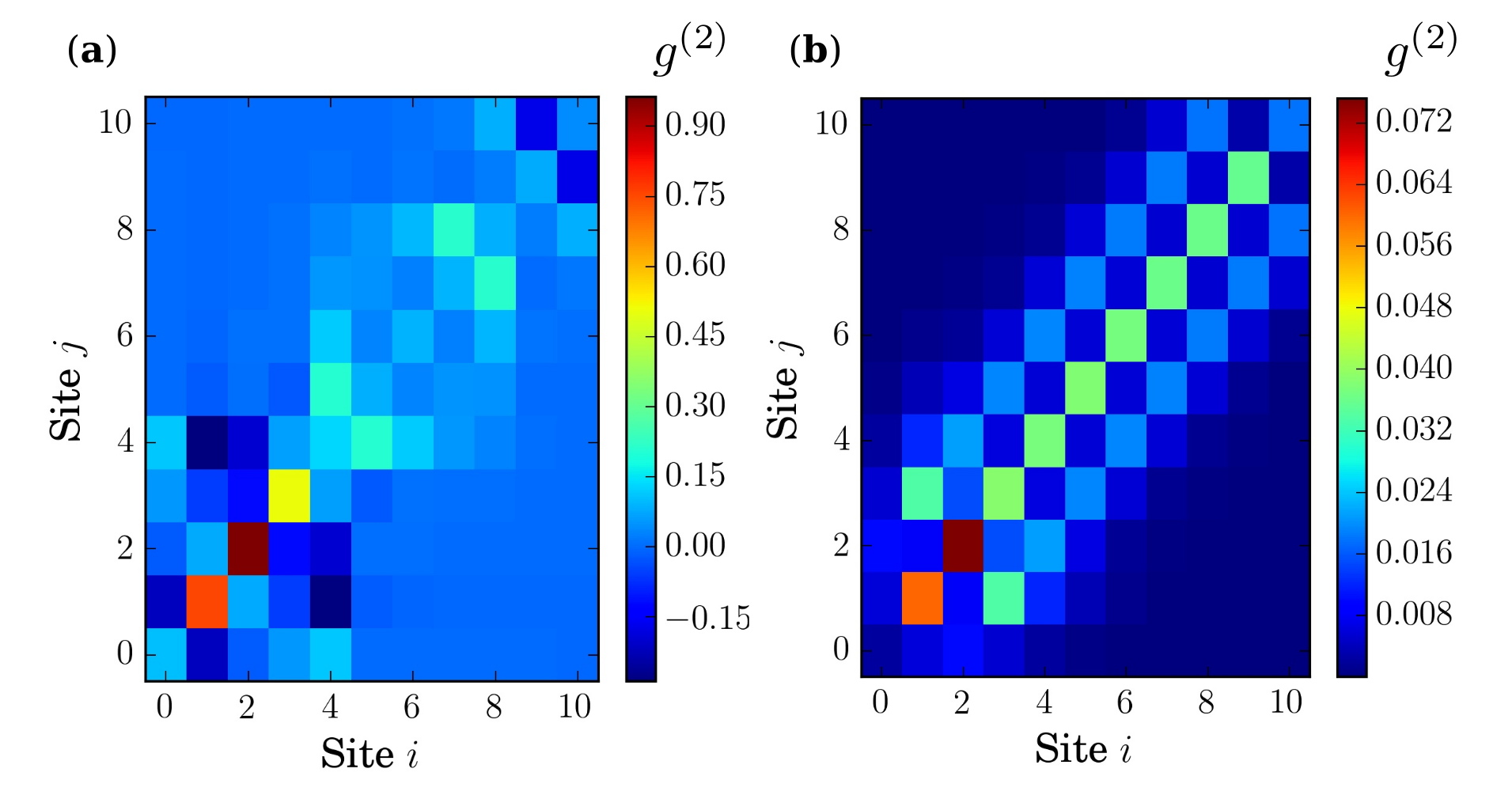}
\caption[Two-point correlators identify superfluid occupation.]{\label{fig:switchG2}\textit{Two-point correlators identify superfluid occupation.} $g^{(2)}$ as a correlation measure distinguishes the presence of superfluid fragments, thus granting an experimentally viable many-body probe of Mott switch states. Both panels illustrate a barrier height tuned to the locality of sharp contrast at $V_0=5$. While the switch flips states between $U=25$ and $U=26$, this level of precision is not necessary for operation. (\textbf{a}) At $U=20$, and time $t=0.05$ for example, the large amplitude of $g^{(2)}$ from $0.9$ to $-0.15$ conveys a level of superfluidity in the disconnected switch state, which is more than an order of magnitude larger than (\textbf{b}) the connected switch when $U=30$ also at time $t=0.05$, which ranges from $g^{(2)}=0.072$ to $0.008$.}
\end{figure}

Furthermore, the $g^{(2)}$ measure delineates second order correlations that not only provide ease of experimental access and drive local entropy, they act as a guide for entanglement, if not a witness~\cite{Song2010}. These $g^{(2)}$ correlators gauge particle fluctuations for optical lattice sites $i$ and $j$ and so $g^{(2)}_{ij} = \langle \hat{n}_i \hat{n}_j \rangle - \langle \hat{n}_i \rangle \langle \hat{n}_j \rangle$. When $g^{(2)}$ is positive, the expectation value of observing two atoms concurrently at sites $i$ and $j$ is greater than that of observing the distinct atoms locally and vice versa. Because the correlations are positive across most of the lattice, this corresponds with a preference for simultaneous two-body measurement over local measurement. Figure~\ref{fig:switchG2}(\textbf{a}) portrays the initial (time $t=0.05$) two-point correlator of the switch in its disconnected state for $U=20$; the small amount of superfluid on the Mott background resulting from the confinement of the barrier magnifies the $g^{(2)}$ amplitude by an order of magnitude over Figure~\ref{fig:switchG2}(\textbf{b}), which illustrates initial ($t=0.05$) normal switch operation for $U=30$. This normal switch state is also marked by the extreme localization of the two-point correlations on-site and of nearest neighboring lattice sites.

\section{Conclusions}
We have demonstrated a proof-of-principle of a robust atomtronic switch in a double well Mott insulator by triggering a local excitation in the Mott background and measuring a transient response. The switching mechanism is tuned via the lattice depth and thus the local interaction strength triggers a sharp phase transition. The switching occurs due to the confinement of the double well barrier, which modifies the traditional superfluid-Mott ground states. In the disconnected state, the excitation is a particle-hole pair; in the connected state, the excitation is a hole -- an absence of a particle -- within the Mott insulating background. The geometry thus changes between the two states, creating a critical transition in the conductance, a point that can be quantified via time-dependent Fock state transmission of the hole through the barrier, which also provides a means of determining signal-to-noise ratio and operating regime. We further demonstrate the fidelity of the Fock states and the corresponding discrete Fourier transform as a tool for optimizing and characterizing the switch noise performance. What is more, $g^{(2)}$ as a measure of the two-point correlations provides a witness of superfluid fragments on the Mott background, thus demonstrating a many-body probe of Mott switch states that is feasible on current cold atom platforms. The phase transition modulated switch mechanism proposed in this article offers flexibility due to large parameter margins for operation together with sharp contrast of the switch states. 

The next steps of many-body quantum simulation of robust Mott atomtronics switches, particularly their applications in strongly-correlated regimes, would address problems of materials science or other disciplines more directly, even in combination with open quantum systems for improved source and drain implementation. We could also look toward ultracold molecules for additional degrees of freedom, where the transport is of a spin rather than a mass~\cite{lusk2015,zang2017}. Then we get a ``moleculetronics'' switch which is in fact spintronics in the ultracold context.

\section{Acknowledgements}
The authors would like to extend gratitude to Diego Alcala for extensive support during the conception of this project -- and to Elias Galan and Lewis Graninger for making this paper possible. Many heartfelt thanks to Steven Patton for suggesting the original idea for the project. High performance computing resources at the Colorado School of Mines and the Golden Energy Computing Organization were used to perform simulations. This research is partially based on work supported by the US National Science Foundation under grant numbers PHY-1520915, PHY-1207881, PHY-1306638, OAC-1740130, as well as the US Air Force Office of Scientific Research grant number FA9550-14-1-0287. We acknowledge the support of the U.K. Engineering and Physical Sciences Research Council (EPSRC) under the ``Quantum Science with Ultracold Molecules'' Programme grant number EP/P01058X/1.

\bibliographystyle{plainurl}
\bibliography{refs}

\section{Appendix}
\begin{figure}
\centering
\includegraphics[width=0.65\textwidth]{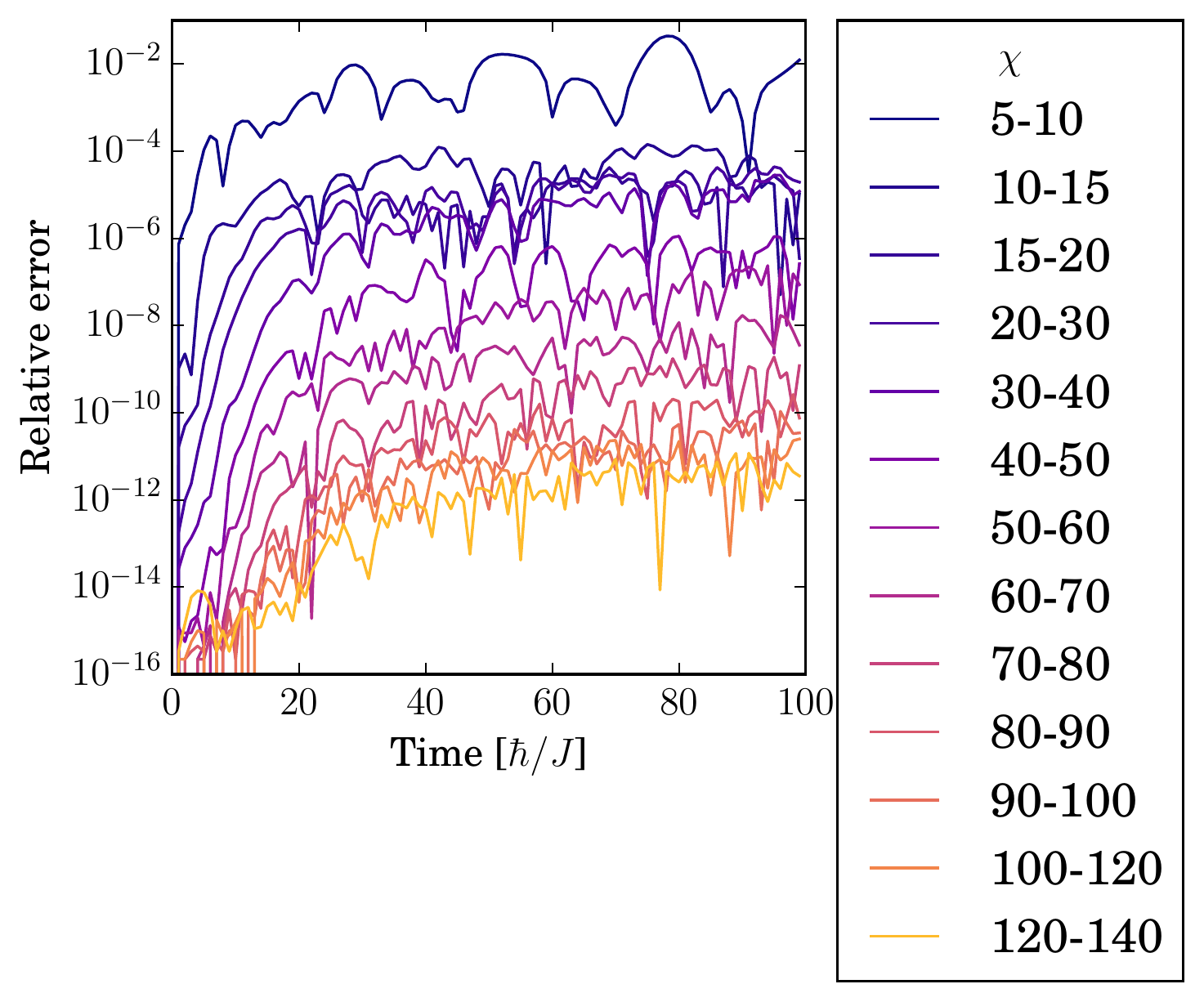}
\caption[Convergence of the bond dimension.]{\label{fig:convergence}\textit{Convergence of the bond dimension.} In the proposed switch operating regime for $L=11$, $N=10$, $U=30$ and a barrier height $V_0=5$, we demonstrate convergence below $10^{-11}$ of the Fock state fidelity for bond dimension up to $\chi=140$.}
\end{figure}

Time-evolving block decimation errors can be split into two categories, $\epsilon = \epsilon_{\mathrm{method}}+\epsilon_{\chi}$, where $\epsilon_{\mathrm{method}}$ is from a number of sources; the most significant of these are from the fifth-order Suzuki-Trotter approximation that induces a time step error, together with the local dimension truncation. $\epsilon_{\chi}$ dictates the Schmidt truncation error due to truncation of the Hilbert space. We found the time step in the limit of small $U$ and $V_0$ to require excessively small time steps to converge; whereas in the large $U$ and $V_0$ limit, the bond dimension $\chi$ required for convergence was excessively large. In the parameter space relevant to this paper, $U=20$ through $30$ and $V_0=0.1$ through $10$, the $V_0=0.1$ case is converged below $10^{-4}$ relative error in the bond dimension, and the $V_0=10$ case is also converged below $10^{-4}$ due to limitations in the bond dimension. All other barrier heights for any interaction strength in this range were converged below $10^{-7}$, though we check bond dimension up to $\chi=140$ in our preferred parameter regime, where the convergence is actually orders of magnitude better, as shown in Figure~\ref{fig:convergence}. Other interaction ranges were also considered as part of this project, from $U=4$ to $60$, where the regions from $U=31$ to $60$ and $U=4$ to $19$ were converged below $10^{-3}$ for convenience.

\end{document}